\magnification=\magstep1
\baselineskip=17 pt plus 2pt minus 2pt
\vsize=9.2truein
\hsize=7.0truein
\hoffset=-0.3truein
\hfuzz=8pt
\footline={\ifnum\pageno=1 \hfil\else\centerline{\folio}\fi}
\font\chapter=cmbx12 scaled \magstep 1
\font\smchapt=cmbx12

\font\smaller=cmr9
\font\reglar=cmr10
\font\trila=cmmib10 \textfont9=\trila
\mathchardef\beta="710C \mathchardef\gamma="710D
\mathchardef\delta="710E \mathchardef\lambda="7115
\mathchardef\russ="715A
\font\hscript=eufb10 scaled \magstep 1  
\def\hsp{{\hscript h}}

\textfont9=\trila

\def\hi{\hangindent=15pt}
\def\vs{\vskip 6truept}

\def\no{\noindent}
\def\sp#1{\vs\noindent{\bf \llap{#1.~~}}}

\def\p{\partial}
\def\bq{\tilde q}
\def\bp{\tilde p}

\def\E#1{Eqs.~(#1)}

\def\bgamma{{\fam=9 \gamma}}

\def\tfrac#1#2{{\textstyle{#1\over#2}}}
\def\dfrac#1#2{{\displaystyle{#1\over#2}}}
\def\sfrac#1#2{{\scriptstyle{#1\over#2}}}

\def\rep{\quad{=\llap{$\longrightarrow$}\quad}}
\def\Long{{{\quad\Longrightarrow\quad}}}
\def\E#1{Eq.~(#1)}
\def\Es#1{Eqs.~(#1)}
\def\overbar#1{{\overline{#1}}}
\def\dom{{\dform{\omega}}}

\def\gam{{\dform{\Gamma}{}}}
\def\OM{{\dform{\Omega}{}}}

\def\sn{\mathop{{\rm sn}}\nolimits}        
\def\cn{\mathop{\rm cn}\nolimits}
\def\dn{\mathop{\rm dn}\nolimits}

\def\dform#1{\smash{\raise.5pt\hbox{$\mathop{#1}\limits_{
    \raise3pt\hbox{\smash{\hbox{$\scriptstyle\sim$}}}}$}}}

\def\tc#1#2#3#4{{{{\vartheta\raise1.8pt\hbox{${\scriptstyle{\left[{#1\atop#2}\right]}}$}
            (#3|\,#4)}}}}
\def\ch#1#2{{{{\vartheta\raise1.8pt\hbox{${\scriptstyle{\left[{#1\atop#2}\right]}}$}}}}}
\def\vt#1#2{{{\vartheta_{#1}(#2|\tau)}}}

\def\sig{{{\dform{\sigma}}}}
\def\ftp#1{${}^{{\scriptscriptstyle #1}}$}
\def\gee{{{\dform{G}}}}

\def\cw#1{{{\cal W}_{\pm #1}}}
\def\vt{{{\vartheta}}}
\def\cdf{{\dform{\cal C}}}
\def\starthree{{{\mathop{*}_\gamma}}}
\def\Ep{{{{\dform e}_+}}}
\def\Em{{{{\dform e}_-}}}

\def\of{{{\overline f}}}

\def\ovq{{{{\widetilde {Q}}}}}
\def\vq{{{ Q}}}
\def\sq{{{2}}}   

\reglar
{\smchapt \centerline{ Solutions of the sDiff(2)Toda equation with SU(2) Symmetry}}

\vs
\centerline{Daniel Finley$^\dagger$ and John K. McIver$^\ddagger$}\par\no
\centerline{$^\dagger$University of New Mexico, Albuquerque, NM 87131, USA}\par\no
\centerline{$^\ddagger$University of Idaho, Moscow, ID 83843, USA}
\vskip12pt
{\narrower{\narrower
\par\no
{\bf Abstract}\par\no
We present the general solution to the Pleba\'nski equation for an \hsp ~space that admits Killing
vectors for an entire SU(2) of symmetries, which is therefore also the general solution of the sDiff(2)Toda
equation that allows these symmetries.  Desiring these solutions as a bridge toward the future for
yet more general
solutions of the sDiff(2)Toda equation, we generalize the earlier work of Olivier, on the Atiyah-Hitchin
metric, and re-formulate work of Babich and Korotkin, and Tod, on the Bianchi IX approach to a metric with an SU(2)
of symmetries.  We also give careful delineations of the conformal transformations required to ensure that
a metric of Bianchi IX type has zero Ricci tensor, so that it is a self-dual, vacuum solution of the
complex-valued version of Einstein's equations, as appropriate for the original Pleba\'nski equation.
\vskip12pt\no
PACS numbers:\hskip0.4in 04.20.Jb
\vs}\vs}
\vs
\sp{I}{\bf Introduction}
\vs

We have long been interested in the Pleba\'nski\ftp1 formulation for an \hsp-space, i.e.,
a 4-dimensional, complex manifold with
an anti-self-dual conformal curvature tensor\ftp2 that also satisfies the Einstein vacuum field equations.
Any such space is determined by a solution to the Pleba\'nski heavenly equation, a constraining
pde for a single function of 4 variables.  By now many different, equivalent forms of that equation
have been developed;
however, the two most common are the original ones given by Pleba\'nski:  the
first form, for a function $u =  u(p,\bp, q,\bq)$,
and the second form, for a function $v=v(x,y,p,q)$, either of which serve as potentials for the metric
via their second derivatives:
$$\eqalign{u_{,p\bp}u_{,q\bq} - u_{,p\bq}u_{,q\bp}
= 1&\hbox{~~and~~} {\bf g} = 2(u_{,p\bp}\,dp\,d\bp + u_{,p\bq}\,dp\,d\bq + u_{,q\bp}\,dq\,d\bp +
 u_{,q\bq}\,dq\,d\bq)\,,\cr
 &\hbox{or~~}\hskip3.7in\llap{(1.1)}\cr
v_{,xx}v_{,yy}-v_{,xy}^2 + v_{,xp}+v_{,yq} = 0&
\hbox{~~and~~}
{\bf g} =  2dp(dx-v_{,yy}dp+v_{,xy}dq) +  2ds(dy+v_{,xy}dp - v_{,xx}dq)\;,}$$
 where partial derivatives are
indicated by a subscript which begins with a comma.
This approach already has a long history; nonetheless, there still seems to be considerable
effort being made\ftp3 to better understand the structure of the space of solutions, and any notion as
to the behavior of the ``general" solution is still far from being found.
\vs
Nonetheless, to make some progress on that problem, one looks to simplify the question.
A standard approach is to simplify the question by looking for
those metrics that admit a Killing vector.
If the (necessarily skew-symmetric) covariant derivative of that Killing vector has
an anti-self-dual part that vanishes, then the problem has indeed been completely resolved,\ftp{4} filling
in some part of the solution space.
In this case the constraining equation for $u$ can be reduced to simply an appropriate solution of
 the 3-dimensional Laplace equation.\ftp4
However, when that anti-self-dual part is non-zero, the Killing vector reduction instead gives
 a single pde for an unknown function $\Omega=\Omega(q,\bq,s)$ of only
three remaining (complex) variables, which we refer to as the sDiff(2) Toda equation\ftp5 because of its
symmetry properties.  The equation, again along with the form of the
metric\ftp7 that it generates is the following,
 where we use $\varphi$ as the coordinate along the flows
generated by the Killing vector:
$$\eqalign{\Omega_{,q\bq} + \left(e^\Omega\right)_{,ss} = 0 \;,&\qquad
{\bf g} =  V\,{\bgamma} + V^{-1}
(d\varphi + \dform{\omega})^2\;,\cr
V \equiv \tfrac12 \Omega_{,s}\;,\quad& \bgamma \equiv ds^2 + 4\,e^\Omega\,
dq\, d\bq\;,\qquad\dform{\omega} \equiv \tfrac{i}2\{\Omega_{,q}dq - \Omega_{,\bq}
d\bq\}\cr}\eqno(1.2)$$
This equation has been of interest in general relativity in various contexts, as well as some
other fields of theoretical physics, for over twenty years.  Many explicit solutions have been
found\ftp8 although few solutions of general type are known, especially for cases without any additional
Killing vectors.  Hoping to understand methods to find solutions without additional Killing vectors,
there have,
fairly recently, been several deliberate searches
for so-called ``non-invariant" solutions of the sDiff(2) Toda equation.\ftp9
\vs
However, it is true that the use of symmetries continues to be the most efficient method we have for
solving nonlinear pde's.  Fairly early in the studies of this equation,  M.A. Saveliev\ftp{10} used it
as a platform to begin his study of continuum Lie algebras,\ftp{11} and even presented a form which,
in principle, gave the desired ``general solution" of the equation in terms of some initial conditions.
Unfortunately that form is very complicated and does not seem to be useful for obtaining
manageable and interesting, specific new solutions.  Various explications of the symmetries, and the
generalized symmetries have been made.\ftp{12}  It is hoped that the detailed characterization
of those symmetries facilitates both their use for finding additional classes of general solutions
as well as checking for noninvariant solutions.  However, we propose a different approach in this article.
\vs
There is one particular subset of all solutions that has truly received extensive study
during the last 15 or so years; this is when the metric allows
an entire SU(2) [or, equivalently, an SO(3)] of symmetries.\ftp{13}  In this case
Einstein's field equations are reduced to simply a system of
ordinary differential equations, for functions of one remaining independent variable.
  At least the question of the solutions themselves
has been completely answered for this case.  On the other hand, the
usual approach to this problem begins from an entirely different mechanism, built specifically on that large
family of symmetries, and usually referred to as a solution of Bianchi type IX.  Those solutions are
formulated either via Schwarzian triangle functions\ftp{14} or Painlev\'e transcendents.\ftp{15}
Because the mechanism used to obtain these solutions is quite different from one that would begin
 from Pleba\'nski's
equation it is not of as much use as we
would like, from the point of view of using it
 as a starting point to work backwards and find more general classes of solutions
of the sDiff(2) Toda
equation.  Indeed other researchers have also thought about this question, and work of
Olivier\ftp{16} and also of Tod\ftp{17}
has shown many of the details of relationships of some of the cases
with solutions of the sDiff(2) Toda equation; however,
the essential purpose of this paper is to make more explicit that sort of information about functions
$\Omega(q,\bq,s)$, constituting solutions of our equation, with the hope that this will help push
forward that search for a much better understanding of the solution space.  In particular we will
begin with the most general form for a vacuum, Bianchi IX solution, in the form given by Babich and
Korotkin,\ftp{18} who express their solutions in terms of elliptic theta functions.  Then we will
show how one may determine the corresponding variables for the sDiff(2) Toda equation.  It is perhaps also
worth noting here that other approaches to extending these solutions, by Iona\c{s},\ftp{19} and by Ohyama,\ftp{20}
have emphasized the desirability of using elliptic theta functions for problems related to this.
\vskip12pt

\sp{II}{\bf The Bianchi IX Approach}
\vs
We begin  here with the important details of the formalism usually used for investigations
of Bianchi IX metrics, using notation that follows Babich and Korotkin.\ftp{18}
  The assumed three Killing vectors give
foliations of the 4-space, as orbits of those Killing vectors,
 that are, at least topologically, spheres, so that a very
reasonable approach is to use the Pauli 1-forms, $\{\sig_i\}_1^3$,
 as a Maurer-Cartan basis, for the assumed SU(2) symmetries.
 In addition we use a fourth coordinate normal to those surfaces, that we denote by $\mu$.
The general form of the Bianchi IX metric, with additional conformal factor, which may be resolved so
that it has zero Ricci tensor and also {\bf has} a conformal tensor which is either anti-self-dual
or self-dual is usually stated as follows:
$$\eqalign{
{\bf g} = F^2w_1w_2w_3\,&\left\{d\mu^2 + {\sig_1^2\over w_1^2} + {\sig_2^2\over w_2^2}
 + {\sig_3^2\over w_3^2}\right\}\
 \;,\cr
 d\sig_1 = &\sig_2\wedge\sig_3\;,\quad  d\sig_2 = \sig_3\wedge\sig_1\;,\quad
d\sig_3 = \sig_1\wedge\sig_2\;,\cr
}\eqno(2.1)$$
where the three functions
$\{w_i\}_1^3$ and the conformal factor $F$ depend only on $\mu$.
The requirements concerning  anti-self-duality or self-duality of the conformal tensor,
and the vanishing of the Ricci scalar,
are turned into (a system of ordinary) differential equations that must be satisfied by the three functions
$\{w_i\mid i=1,2,3\}$.
The vanishing of the quantities labeled as ${\cal W}_{+i}$ below cause the self-dual part of the conformal
tensor to vanish, causing the resultant space to have an anti-self-dual conformal tensor, while the
vanishing of the other set, ${\cal W}_{-i}$ cause the anti-self-dual part to vanish, with the opposite
conclusion, where we do notice that the one version may be changed into the other simply by a change of
the sign of the independent variable, $\mu$, its derivative being indicated below simply by a prime:
$$\left.\eqalign{\cw{i}\ \equiv &
\mp {a_{\pm i}}^\prime + a_{\pm i}(a_{\pm j}+a_{\pm k}) - a_{\pm j}a_{\pm k}\;,\cr
2a_{\pm i} = Y_{\pm j}+Y_{\pm k}-Y_{\pm i}\;,&\quad
Y_{\pm i}\equiv \pm{w_i'\over w_i} + {w_jw_k\over w_i}\equiv a_{\pm j}+a_{\pm k}\;,\cr}\right\}\
i,j,k = 1,2,3,\hbox{~cyclic,}\eqno(2.2)$$
where we use
the word ``cyclic" with a fairly standard meaning, i.e.,
to mean that the indices $\{i,j,k\}$ should always take distinct
values and in cyclic order, so that they imply each of the three possible sets of values
$1,2,3$,  and $2,3,1$, and $3,1,2$.
A complete derivation of the provenance of these equations is given in Appendix I.
\vs
Since we have chosen, following historical precedent with the Pleba\'nski approach to these problems, to
concern ourselves with the anti-self-dual solution, we will ask that the ${\cal W}_{+i}$ should vanish.
With that choice we will write out more explicitly the coupled set of six
equations which must be solved, and will suppress
the $+$-subscript on what was called $a_{+i}$ above since this is the only one with which we will be
concerned:
$$\left.\eqalign{0 = {\cal W}_{+i}\Long &a_i^\prime + a_ja_k = a_i(a_j+a_k)\;,\cr
&w_i^\prime + w_jw_k = w_i(a_j+a_k)\;,\cr}\right\}\ i,j,k = 1,2,3,\hbox{~cyclic.}\eqno(2.3)$$
The general case of a solution of these equations, with no constraint on the conformal factor, $F=F(\mu)$,
will have a non-zero, although traceless Ricci tensor.  The equations for the three functions $\{a_i\}_{i=1}^3$
are often referred to as the Halphen system,\ftp{21} and
there are several different known forms for the solution,
involving Painlev\'e transcendents of type III,\ftp{17}, Schwarzian triangle functions,\ftp{14}
 or complete elliptic integrals.\ftp{22}
We will not often need to use the explicit solutions; nonetheless, we prefer the form given by
Babich and Korotkin,\ftp{18} in terms of {\it theta functions}.  The general solution for the three
functions $\{a_i\mid i=1,2,3\}$ is a 3-parameter one given in terms of the general M\"obius transformation
of the upper-half of the complex $\tau$-plane when it acts on the following
(generic) particular solution of the system. We replace the usual variable $\tau$ for the theta
functions, which must have positive imaginary part, by $\tau\equiv i\mu$,
so that we may treat $\mu$ as real-valued and positive, befitting its role in the form of our
metric.  The particular solution of the system is given by:
$$\eqalign{ a_{i} = &\ 2{d\over d\mu}\log\vartheta_{5-i}\;,\ i = 1, 2, 3\;;\quad\hbox{where~~}\left\{\eqalign{
\vartheta_2 \equiv &\ \tc{1/2}0{0\,}{\!i\mu} = e^{-\sfrac\pi4\mu}\!\!\!\sum_{m=-\infty}^{+\infty}e^{-\pi m(m+1)\mu}\;,\cr
\vartheta_3 \equiv &\ \tc{0}0{0\,}{\!i\mu} = \sum_{m=-\infty}^{+\infty}e^{-\pi m^2\mu}\;,\cr
\vartheta_4 \equiv &\ \tc0{1/2}{0\,}{\!i\mu} = \sum_{m=-\infty}^{+\infty}(-1)^m\,e^{-\pi m^2\mu}\;,\cr
}\right.\cr}\eqno(2.4)$$
The M\"obius
transformation sends $\tau \rightarrow (a\tau + b)/(c\tau + d)$, with $ad-bc = 1$.
 More details of the transformation and how it scales the functions $a_i$,
are given in Appendix III.  As well we describe the general theta functions, which also depend on a complex
variable $z$.  When those functions are evaluated at $z=0$, as they are above, they are often referred to as {\it
theta coefficients}, and are related to the complete elliptic integrals, $K(k)$ and $E(k)$.
\vs
The functions $w_i(\mu)$ are related to these others.  For the case when the Ricci scalar vanishes
they are related in a very simple way, involving the conformal factor.
We are interested only in the pure
vacuum case, so that the entire Ricci tensor vanishes.  We will
explain in detail in Appendix I that this can be done provided one chooses
$$F(\mu) = c_0(\mu + d_0)\,,\hbox{~and~~} w_i = a_i + {d\over d\mu}\log F\;,\eqno(2.5)$$
 for arbitrary constant values $c_0$ and $d_0$
as discussed in the paper of Babich and
Korotkin,\ftp{18} with reference to work of Tod.\ftp{25}  It is also worth noting that the first solution
of this sort, determined by Atiyah and Hitchin\ftp{22}, and explained more fully by Olivier\ftp{16} as regards
its relation to Pleba\'nski's
equation, corresponds to the limit where $c_0\rightarrow 0$ at the same that $c_0\,d_0\rightarrow$ a non-zero
constant, corresponding to
a constant value for $F$, and therefore $w_i = a_i$.
\vskip15pt
\sp{III}{\bf The Matching Process}
\vs
To create the desired mappings between the two sets of variables, we will need parametrizations of
the Pauli 1-forms, using a set of Euler angles on the sphere, and identified in such a way that
 the coordinate angle $\varphi$ represents the variable along our original
 Killing vector, the one required by the sDiff(2)Toda equation itself.  We choose one
such representation given by Tod:\ftp{17}
$$\eqalign{
\sig_1+i\sig_2 =e^{i\psi}(d\theta - i\sin\theta\,d\varphi)\;,\quad
\sig_3 = d\psi + \cos\theta\,d\varphi\;.}\eqno(3.1)$$
We begin the process of comparing the two forms of the metric, and establishing coordinate transformations by
first considering that the terms involving $d\varphi^2$ are uniquely picked out, since it is our ``obvious"
Killing vector.  Expanding out the metric in terms of these Euler angles,
we have the following identification, as a first step:
$$\eqalign{\dfrac1V = |\hskip-1pt|\p/\p\varphi|\hskip-1pt|^2 &\ =
 F^2\{[{w_1w_3\over w_2}\cos^2\psi+{w_2w_3\over w_1}\sin^2\psi)\sin^2\theta +
{w_1w_2\over w_3}\cos^2\theta\}\equiv {F^2\over w_1w_2w_3}M\;,\cr
&M \equiv \left[w_3^2(w_2^2\sin^2\psi + w_1^2\cos^2\psi)\sin^2\theta + w_1^2w_2^2\cos^2\theta
\right]\;,}\eqno(3.2)$$
where we have defined the quantity $M$ for convenience since it will appear in many different places in the
discussion.  As the 3-dimensional portion of our metric, $\bgamma$, does not depend on $\varphi$, we may next identify the
1-form $\dform\omega$ in the Pleba\'nski form of the metric:
$${1\over V}\,{\dform\omega} = F^2\Big\{\left({w_2w_3\over w_1} - {w_3w_1\over w_2}\right)\sin\theta\sin\psi\cos\psi
d\theta
+ {w_1w_2\over w_3}\cos\theta\,d\psi\Big\}\eqno(3.3)$$
\vs
With these forms  we may now calculate  $\tfrac1V(d\varphi + {\dform\omega})^2$ and remove that term
from the Bianchi IX form of the metric, providing us with
the form of $\bgamma$ in the those coordinates:
$$\eqalign{\bgamma& = F^4\Big\{M\,d\mu^2 +[w_3^2\sin^2\theta +
(w_1^2\sin^2\psi+w_2^2\cos^2\psi)\cos^2\theta]\,d\theta^2 \cr
&\qquad+2(w_1^2-w_2^2)\sin\psi\cos\psi\sin\theta\cos\theta\,d\theta\,d\psi
+ (w_1^2\cos^2\psi+w_2^2\sin^2\psi)\sin^2\theta\,d\psi^2\Big\}\cr
&\qquad\qquad = \gamma_{\mu\mu}d\mu^2 + \gamma_{\theta\theta}d\theta^2 + 2\gamma_{\theta\psi}d\theta\,d\psi
+ \gamma_{\psi\psi}d\psi^2 =  ds^2 + 4\,e^\Omega\,dq\, d\bq\;.}\eqno(3.4)$$
Our next step is to determine the factor $ds^2$, which is related to the 1-form ${\dform\omega}$ as follows,
beginning from \Es{1.2}:

$$\eqalign{2V = &\Omega_{,s}\,, \quad\dform{\omega} \equiv \tfrac{i}2\{\Omega_{,q}dq - \Omega_{,\bq}
d\bq\}\ \Longrightarrow\ -i\,d{\dform\omega} = V_{,q}ds\wedge dq - V_{,\bq} ds\wedge d\bq + \Omega_{,q\bq} d\bq\wedge dq\;,\cr
\Rightarrow\ \starthree& d{\dform\omega} = {(e^\Omega)_{,ss}\over e^\Omega}ds + V_{,q}dq + V_{,\bq}d\bq
=  V_{,q}dq + V_{,\bq}d\bq +(V_{,s}+2V^2)ds = V^2d(2s-1/V)\;,}
\eqno(3.5)$$
where we have used the fact that $\Omega$ must satisfy its constraining pde, and we note that the Hodge dual in question
is the one generated by the 3-metric $\bgamma$ in the Pleba\'nski form, which says
$$\starthree (dq\wedge ds) = i dq \;,\quad \starthree(ds\wedge d\bq)= i d\bq \;,\quad
\starthree(dq\wedge d\bq) = -\tfrac{i}2 e^{-\Omega} ds\,\;.\eqno(3.6)$$
As we have $\dform\omega$ and $1/V$ in terms of the Euler angles, we may solve this for the desired $d(2s)$:
$$d(2s) = V^{-2}\starthree{\dform\omega} + d(1/V) =  d(1/V) +
\starthree\left[{1\over V}d\left({1\over V}{\dform\omega}\right)
+ {1\over V}{\dform\omega}\wedge d{1\overwithdelims()V}\right]\;,\eqno(3.7)$$
where it is re-expressed in this second form since it then involves only those quantities that we already know.
Of course, in order to do that we
must first determine the dual mapping, but now relative to $\bgamma$, in terms of these
coordinates---rather more complicated than the other ones because of its off-diagonal
terms.   We give the results below of the duals
 of the three basis 2-forms, sufficient to determine what is wanted.
  This calculation is straightforward, if
rather lengthy, and is described in more detail in Appendix II, with the following result:
$$\eqalign{ds = &\ F^2\Big\{[w_3(w_1\cos^2\psi + w_2\sin^2\psi)\sin^2\theta + w_1w_2\cos^2\theta]d\mu\cr &\quad
+ (w_3 - w_1\sin^2\psi - w_2\cos^2\psi)\sin\theta\cos\theta\,d\theta
+ (w_2-w_1)\sin\psi\cos\psi\sin^2\theta\,d\psi\Big\}\cr
& \equiv {\cal F}\,d\mu + {\cal G}\,d\theta + {\cal H}\,d\psi\;.}
\eqno(3.8)$$
We need to square this and subtract appropriately from $\bgamma$ so as to determine the remainder of the transformation
equations, which will be done shortly.  Nonetheless, we will first note here that this equation can
 be explicitly integrated, to give this Pleba\'nski coordinate as a function of the ones used in the Bianchi IX approach:
$$s=s(\mu,\theta,\psi) = \tfrac12F^2[w_1+w_2+(w_3 - w_1\sin^2\psi - w_2\cos^2\psi)\sin^2\theta]\;.\eqno(3.9) $$
\vs
Returning now to the forms given in \Es{3.4}, we want to have $\bgamma - ds^2$ in the form
$e^\Omega(2dq)(2d\bq)$; i.e.,  we
want a pair of coordinates $q$ and $\bq$, and a function $\Omega$ such that this equation would be satisfied.
However, what we actually have is the
difference of those two second-rank tensors, which we want to describe in the form above.  Therefore, it is first
useful to factor that difference into a pair of (complex-valued) 1-forms, which we call
 ${\dform{e}}_+$ and its ``conjugate," $\Em$, such that
$$\Ep\Em \equiv \bgamma- (ds)^2\;.\eqno(3.10)$$
We may then find a function $e^\Omega$ so that the separate 1-forms that make up that product
may each be integrated.  Therefore, we
define a complex-valued scalar function, $f=f(\mu,\theta,\psi)$, to be chosen so that
$$\left.\eqalign{d( e^{-f}\Ep) =&\ 0\;,\cr d(e^{-\overbar f}\Em) = &\ 0\;,\cr}\right\}
\quad e^{f}e^{\overbar f} = e^\Omega\;,\Long \left\{\eqalign{\Ep = e^f\,2dq\;,\cr \Em = e^{\overbar f}\,2d\bq
\;.}\right.\eqno(3.11)$$
where $q$ and $\bq$, are the desired new coordinates.
We begin with the desired factorization into the product of two 1-forms:
$$\eqalign{\Ep = &F^2\sin(\theta)\left\{
\left[w_3(w_1\cos^2\psi + w_2\sin^2\psi)-w_1w_2\right]\cos\theta
 + iw_3(w_2-w_1)\sin\psi\cos\psi\right\}\,d\mu\cr
 & - F^2\left\{(w_1\sin^2\psi+w_2\cos^2\psi)\cos^2\theta + w_3\sin^2\theta -i(w_2-w_1)\sin\psi\cos\psi\cos\theta
 \right\}\,d\theta\cr
 &\qquad +F^2\sin\theta\left\{(w_2-w_1)\sin\psi\cos\psi\cos\theta - i(w_1\cos^2\psi + w_2\sin^2\psi)\right\}
 \,d\psi\cr
 &\equiv {\cal A}\,d\mu + {\cal T}\,d\theta + {\cal L}\,d\psi\;,}\eqno(3.12)$$
while the other one, $\Em$, is obtained from $\Ep$ simply by changing all the $i$'s above to $-i$'s, i.e.,
treating it as if all our variables are real-valued and they should be complex conjugates of each other.
It is a lengthy but straightforward calculation to verify that this
pair does indeed accomplish the desired factorization scheme.  The necessary
``integrating factor," i.e. the function $e^f$, is not uniquely determined; nonetheless the
choice that we find acceptable is
 given, conveniently, in terms of its square as follows:
$$ e^{2f} \equiv \sq F^2\left[(w_1-w_2)(\sin\psi + i\cos\psi\cos\theta)^2 + (w_3-w_1)\sin^2\theta\right]\;,\eqno(3.13)$$
with the conjugate $e^{2{\overbar f}}$ again obtained simply by changing the $i$ above to $-i$.  It is true that this
is the square of the desired factor, rather than the factor itself;
however, because of its complex nature it is better to
display it in this form rather than insisting on just which square root is appropriate. It is then again
straightforward algebra to show the following:
$$e^{3f}d(e^{-f}\Ep) = e^{2f}\,d\Ep - \tfrac12\,de^{2f}\wedge\Ep = 0\;,\eqno(3.14)$$
which {\bf guarantees} the existence of $q$, and also $\bq$, so that we may now write
explicitly their differentials.
\vs
While the necessary forms have indeed already been written out explicitly above, in \Es{3.12},
 the integrations that need to be performed, to determine the explicit form of $q$ and $\bq$,
are much more easily performed in a slightly different set of coordinates.  We will choose a new
set, $\{\mu,\theta,p\}$, replacing $\psi$ in favor of $p\equiv i(\psi+\pi/2)+\log\tan(\theta/2)$, which gives
$dp = i\,d\psi + d\theta/\sin(\theta)$.  This will allow us to determine all the dependence on $\theta$
quite explicitly and easily, leaving differential equations to be solved only in terms of $\mu$ and $p$.
We must then re-express $\Ep$
 in terms of $\{\mu,\theta,p\}$:
$$\eqalign{e^f\,d(2q) &\ = \Ep = {\cal A}\,d\mu + {\cal B}\,d\theta + {\cal C}\,dp\;,\hskip2in u\equiv \cosh p\;;\cr
&{\cal A} \equiv F^2\sin(\theta)\Big\{w_3(w_1-w_2)u\left[u\cos\theta +\sqrt{u^2-1}\right]
 + w_2(w_3-w_1)\cos\theta \Big\}\;,\cr
 &{\cal B} \equiv -F^2\sin^2\theta\left[(w_1-w_2)u^2 +(w_3-w_1)\right]\quad = \quad -\tfrac12e^{2f} \equiv
  -\tfrac12P^2\sin^2\theta\;,\cr
 &{\cal C}\equiv F^2\sin(\theta)\left[(w_1-w_2)u\left(u+\sqrt{u^2-1}\cos\theta\right) - w_1\right]\;,
}\eqno(3.15)$$
where we have given a  simple symbol for the oft-repeated quantity $u\equiv \cosh p$,
and also given a separate name, $P^2(\mu,p)$, to that portion
of $e^{2f}$ independent of $\theta$.
The compatibility of these pde's for $2q$ has already been shown, i.e.,  we know that such a $q$ exists.
The integration procedure is explained in Appendix III, with the following result:
$$ 2q = \tfrac12P\cos\theta + {\bf N}(\mu,p)\;,\quad P\sin\theta\,2dq = \Ep\;,\qquad P^2 = \sq F^2\left[(w_1-w_2)u^2 +(w_3-w_1)\right]
\;,\eqno(3.16)$$
where ${\bf N} = {\bf N}(\mu,p)$ is shown, in that appendix, to have several, equivalent, forms in terms of
elliptic integrals:
$$\eqalign{ \sqrt{\pi}\,{\bf N}(\mu,p) = &\
 c_0\left[ \tfrac12(\mu+d_0){d\over dz}\log\vartheta_4(z\mid i\mu) + \pi z\right]
 \Big|_{z=\sfrac12F(u,k)/K(k)}\cr
&\ =
c_0(\mu+d_0)\left[K(k)E(u,k)-E(k)F(u,k)\right] + \tfrac\pi2c_0{F(u,k)\over K(k)}\;,\quad u\equiv \cosh p\;.}
\eqno(3.17)$$
\vskip12pt
\sp{IV}{\bf Passing to the sDiff(2)Toda equation}
\vs
We have determined the desired coordinate transformation, in the direction
$$\left.\eqalign{s=&\ s(\mu,\theta,\psi)\;,\cr
q=&\ q[\mu,\theta,p(\psi)]\;,\cr
\bq = &\ \bq[\mu,\theta,p(\psi)]\;,\cr}\right\}
\qquad\hbox{and also~~}\Omega = \Omega(\mu,\theta,\psi)
 = \tfrac12\log\left(e^{2f}e^{2{\overbar f}}\right)\;.\eqno(4.1)$$
However, what we wanted was
 $\Omega = \Omega(s,q,\bq)$, but the equations given are (seriously) transcendental, so that
 they cannot be explicitly solved to provide  $\Omega = \Omega(s,q,\bq)$.
Instead we may look at this set of equations as  a parametric approach to that question:~
  all the different functions named in \Es{4.1} give the desired functions in terms of
 ``parameters," $\mu$, $\theta$, and $p$ [or $p(\psi)$].  (We do note that while the integration
 for $q$ was much simpler in terms of the complex-valued variable $p$, the reality conditions for the more
 general process are simpler to follow if we retreat back to the use of $\psi$ instead.)   In order to
 proceed from there, to show that the
 function $\Omega$ determined in this way does actually satisfy the sDiff(2)Toda equation, given in
 \Es{1.2}, we only need to determine the appropriate partial derivatives, as, for instance, in the
 following simple example:
 $${\p\over \p s}\Omega = \left({\p\mu\over \p s}{\p\over \p\mu} + {\p\theta\over \p s}{\p\over \p\theta}
 + {\p \psi\over \p s}{\p\over \p \psi}\right)\Omega\;.\eqno(4.2)$$
 Of course we also do not have the partial derivatives in the directions given just above, such
 as ${\p\mu\over \p s}$.  However, the
 1-forms that form the basis for our 3-space, in \Es{3.12} for $dq$ and $d\bq$ and \E{3.8} for $ds$,
 along with the form for $e^{2f}$ given in
 \E{3.13}, give us the entries for the Jacobian matrix for the coordinate
transformation between these two sets of coordinates, in the opposite direction:
$$ J \equiv {\p (q,\bq,s)\over \p (\mu,\theta,\psi)} = \bordermatrix{&\mu&\theta&\psi\cr
q&\tfrac12e^{-f}{\cal A}&
\tfrac12e^{-f}{\cal T}&\tfrac{1}2e^{-f}{\cal L}\cr
\bq&\tfrac12e^{-{\overline f}}\overline{{\cal A}}&\tfrac12e^{-{\overline f}}\overline{{\cal T}}&
\tfrac{1}2e^{-{\overline f}}\overline{{\cal L}}\cr
s&{\cal F}&{\cal G}&{\cal H}}\;.\eqno(4.3)$$
\vs
The inverted partial derivatives we need are simply the entries of the inverse matrix to this one.  As a
first step toward determining them in some convenient way, we note that it is even hopeful that
 when the rather complicated values for all these
quantities are inserted into this matrix, Maple finds that its determinant is rather simple:
$$\det(J) = -\tfrac{i}{2} e^{-f} e^{-\overline f}\sin\theta F^6\,M = -\tfrac{i}2\,e^{-\Omega}\sqrt{
\det{\bgamma}}\;,\eqno(4.4)$$
where we have used \E{A2.3b} to insert the determinant of the 3-metric.
Then the inverse of the Jacobian matrix itself, determined by the computer algebra program Maple, is given here:
$$J^{-1}~ \equiv {\p(\mu,\theta,\psi)\over \p(q,{\overbar q},s)} =
\bordermatrix{&q&{\overbar q}&s\cr
\mu&e^f\alpha&e^{\overbar f}{\overbar\alpha}&\zeta\cr
\theta &e^f\tau&e^{\overbar f}{\overbar \tau}&\eta\cr
\psi&e^f\lambda&e^{\overbar f}{\overbar\lambda}&\kappa\cr}{1\over M\,F^2}\;,
\eqno(4.5)$$
with their values given below, and we have recalled $M$ from \E{3.2}:

$$\eqalign{M =&\ w_3^2(w_1^2\cos^2\psi+w_2^2\sin^2\psi)\sin^2\theta + w_1^2w_2^2\cos^2\theta\;,\cr
e^{-f}MF^2{\p\mu\over\p q} = \alpha &\ \equiv
\sin\theta\left\{
\Big[w_3(w_1\cos^2\psi + w_2\sin^2\psi)-w_1w_2\right]\cos\theta\cr
&\qquad - iw_3(w_2-w_1)\sin\psi\cos\psi\Big\}\cr
e^{-f}MF^2{\p\theta\over \p q} = \tau&\ \equiv -w_3(w_2^2\sin^2\psi+w_1^2\cos^2\psi)\sin^2\theta
- w_1w_2(w_1\cos^2\psi+w_2\sin^2\psi)\cos^2\theta\cr
&\hskip2in +iw_1w_2(w_2-w_1)\sin\psi\cos\psi\cos\theta\;,\cr
e^{-f}MF^2{\p\psi\over\p q} =\lambda&\ = \Big\{-(w_2-w_1)[w_1w_2\cos^2\theta+w_3(w_1+w_2-w_3)\sin^2\theta]
\sin\psi\cos\psi\cos\theta\cr
&\hskip-0.5in + i[w_1w_2(w_2\cos^2\psi+w_1\sin^2\psi)\cos^2\theta + w_3^2(w_1\cos^2\psi
+w_2\sin^2\psi)\sin^2\theta]\Big\}/\sin\theta\;, \cr
MF^2{\p\mu\over\p s} = \zeta&\ = w_1w_2\cos^2\theta + w_2(w_1\cos^2\psi+w_2\sin^2\psi)\sin^2\theta\;,\cr
MF^2{\p\theta\over \p s} = \eta&\ = [w_2^2(w_1-w_3)\sin^2\psi+w_1^2(w_2-w_3)\cos^2\psi
]\sin\theta\cos\theta\;,\cr
MF^2{\p\psi\over\p s} = \kappa&\ = (w_2-w_1)[w_3^2\sin^2\theta + (w_1w_3+w_2w_3-w_1w_2)\cos^2\theta]
\sin\psi\cos\psi\;,\cr}\eqno(4.6)$$
where the (not-displayed) partial derivatives with respect to $\bq$ are just the complex conjugates of those
with respect to $q$, where all the various functions we have are {\bf considered to be} real.  An interesting
result that comes from this calculation is a close relationship between those partial derivatives that
involve $\mu$, although we have not been able to determine any useful result from it:
$$ MF^2\,{\p\mu\over \p s} = \dfrac1{F^2}{\p s\over \p\mu}\;,\qquad MF^2\,e^{-f}{\p\mu\over \p q} =
{1\over F^2}e^{\overbar f}{\p \bq\over \p\mu}\;,\eqno(4.7)$$
along with the conjugate of the second equation as well.  On the other hand, with these derivatives in
hand from the inversion of the Jacobian matrix, we may now explicitly evaluate $\Omega_{,s}$, which as
noted in \E{1.2} should equal $2V$.  We first define a name, $R$, for the quantity that we actually
have in hand, from \E{3.13} and also \E{3.15}, the product of $e^{2f}$ and $e^{2{\overbar f}}$,
$$\eqalign{ R\equiv &\ e^{2\Omega} =  e^{2f}e^{2{\overbar f}} =
4F^4\Big\{\left[(w_1-w_2)(\sin\psi + i\cos\psi\cos\theta)^2 + (w_3-w_1)\sin^2\theta\right]\cr & \hskip1.4in
\left[(w_1-w_2)(\sin\psi - i\cos\psi\cos\theta)^2 + (w_3-w_1)\sin^2\theta\right]\Big\}\cr
&\hskip1in = \left[2F^2\sin^2\theta\,\left|(w_1-w_2)u^2-(w_1-w_3)\right|\right]^2 = \left[\sin^2\theta
\left|P^2\right|\right]^2\;.\cr}\eqno(4.8a)$$
We may then use the chain rule for $\p/\p s$,
as written out in \E{4.2}, and verify, again via Maple, the (required) relationship that is given in \E{1.2}:
$$
{\p\over \p s}\Omega = \dfrac1{2R}{\p\over \p s}R = 2{w_1w_2w_3\over MF^2} = 2V(\mu,\theta,\psi)
\quad\Long\quad {\p\over \p s}R = 4RV\;. \eqno(4.8b)$$
Next we use these forms to determine a form for one of the desired second derivatives in the sDiff(2) Toda
 equation, namely the second $s$-derivative of $e^{\Omega}$:
$$\eqalign{
& \quad\p_s^2R = 2\p_s(e^\Omega\p_se^\Omega) = 2[e^\Omega\p_s^2e^\Omega + (\p_se^\Omega)^2]\cr
\Longrightarrow  \quad \p_s^2e^\Omega = &\ \dfrac1{2e^\Omega}\left[\p_s^2R - {(\p_sR)^2\over 2R}\right]
= 2e^\Omega V^2\left(2-{\p\over \p s}\dfrac1V\right)\;.}
\eqno(4.9)$$
\vs
It is true that the last formulation above for our second $s$-derivative still has an external factor $e^\Omega$;
however, when we obtain the desired formulation---in terms of the coordinates $\{\mu,\theta,\psi\}$---for the
other second derivative, there will be another such factor, so that they will eventually factor out and not
cause any difficulty.
However, there is the serious difficulty that all of the partial derivatives involving $q$ or $\bq$
contain
$e^f$ and $e^{\overbar f}$, so that it is not immediately obvious how to extract them in the right format.
They can in fact be extracted in the desired form, but we will have to consider not only the product
 $R\equiv (e^{2f})(e^{2\of})$ that we have been considering but also
the quotient, $S\equiv (e^{2f})/(e^{2\of})$, which will allow us to take care of factors of $f$ and $\of$
separately:
$$\eqalign{ &\log R = 2(f+\of)\;,\qquad \log S = 2(f-\of)\;,\cr
\Longrightarrow\quad 2f+\of = &\ \dfrac14(3\log R + \log S)\;,\quad f+2\of = \dfrac14(3\log R-\log S)\;.
}\eqno(4.10)$$
We begin by keeping explicit track of the factors of $e^f$ that appear, writing the following:
$$\p_q =
 {e^f\over MF^2}\left(\alpha\p_\mu +
\tau\p_\theta + \lambda\p_\psi\right)\equiv e^f{ Q}\;,\eqno(4.10)$$
along with its conjugate form for $\p/\p\bq$.
There are then two apparently distinct ways in which the desired second derivative may be
written, which, of course, must be identical since
ordinary partial derivatives commute.  The first ordering is the following:
$$\eqalign{& \p_{\bq}{\p_q}\Omega =  \tfrac12e^\of\,\ovq\left(e^f\vq\log R\right)
= \tfrac12e^\of\,\ovq\left(e^f{\vq R\over R}\right) = \tfrac12e^\of\,\ovq(e^{-f-2\of}\vq R)\cr
&  = \tfrac12e^{-\Omega}\{\ovq\vq R - (\vq R)\ovq(f+2\of)\} = \tfrac12e^\Omega\left\{
{\ovq\vq R\over R} - \tfrac34{\vq R\over R}{\ovq R\over R} + \tfrac14{\vq R\over R}{\ovq S\over S}\right\} \;,
\cr}\eqno(4.11a)$$
which shows that indeed we can re-write our entire equation in such a way as to only need the product $R$,
and the quotient $S$.
The other order for the initial partial derivatives gives the following quantity,
 which appears to be different:
$$\eqalign{\p_q\p_{\bq}\Omega =&\ \tfrac12 \p_q\p_{\bq}\log(e^{2\Omega})
 = \tfrac12e^f{ Q}\left({e^\of}\ovq\log R\right)
= \tfrac12e^f{ Q}\left(e^\of{\ovq R\over R}\right)\cr
& = \tfrac12e^f\vq(e^{-2f-\of}\ovq R) = \tfrac12e^{\Omega}\left\{{\vq\ovq R\over R} -
\tfrac34{\vq R\over R}{\ovq R\over R} - \tfrac14{\vq S\over S}{\ovq R\over R}
\right\}\;.}\eqno(4.11b)$$
\vs
Since these two expressions must actually be the same we may do two useful things with them.  As a first check
on the somewhat complicated algebra, we may first insist that their difference is zero; namely we must have
the following equality, which is simply the statement that the partial derivatives themselves commute:
$$\dfrac4R[\vq,\ovq]R = {\vq S\over S}{\ovq R\over R}
 + {\ovq S\over S}{\vq R\over R}\;.\eqno(4.12)$$
 The calculation (in Maple) involves quite a large number of terms on each side; however, they are in
 fact equal, verifying that all the algebra is correct.  At this point then we finally want to determine
 the desired other second derivative, which takes its most symmetric form via half the sum of the
 two expressions given above in \Es{4.11a-b};
$$\p_q\p_{\bq}\Omega = \tfrac1{4}e^\Omega\,\left\{{\vq\ovq+\ovq\vq)R\over R} -\dfrac3{2}{\ovq R\over R}
{\vq R\over R}
+ \dfrac1{4}{\ovq S\over S}{\vq R\over R} - \dfrac14{\vq S\over S}{\ovq R\over R}\right\}\;.\eqno(4.13)$$
As the expression for $\p^2_se^\Omega$ given in \E{4.9} also has an overall factor of $e^\Omega$, we may
add that expression to this one, and divide out that overall factor, reducing the verification of the
sDiff(2) Toda equation to the question as to whether or not that sum vanishes.  It is a straightforward,
if perhaps somewhat lengthy calculation, performed in Maple, to show that this sum does in fact vanish,
which was the necessary and sufficient condition to guarantee that the parametric presentation we have
obtained is in fact a solution of the sDiff(2) Toda equation.
\vs
\sp{V}{\bf Conclusions}
\vs
The equations we have developed give the general solution to the Pleba\'nski equation,
$\Omega=\Omega(s,q,\bq)$,
 when the manifold is (locally) required to have SU(2) symmetry.  Second derivatives of
 $\Omega$ determine the
components of the metric; therefore $\Omega$ does not depend on the fourth coordinate
 for the manifold, $\varphi$, since variation of it has been chosen for the direction of the
 explicit Killing vector.
  The solution is determined parametrically in
 terms of an additional set of coordinates for the problem, $\{\mu,\theta,\psi\}$,
  so that in fact we have
 our three desired coordinates as functions of them.  While the presentation of
 $\{s,q,\bq\}$ as functions of
 these original coordinates, $\{\mu,\theta,\psi\}$ are explicit in terms of $\theta$
 and $\psi$, we are unable to
 invert those equations explicitly because of the existence within them of the theta
 coefficients, and the
 (elliptic) theta functions, which are
 transcendental functions of $\mu$, analytic for all positive values of $\mu$.
 It is nonetheless useful to
collect those equations together here, from the places where they have been derived in this text:
$$\eqalign{& s = s(\mu,\theta,\psi) = \tfrac12F^2[w_1+w_2+(w_3 - w_1\sin^2\psi - w_2\cos^2\psi)
\sin^2\theta]\;,\cr
&q = q(\mu,\theta,\psi) = \tfrac14P\cos\theta + \tfrac12{\bf N}\;,
\qquad \bq = \bq(\mu,\theta,\psi) = \tfrac14{\overbar P}\cos\theta + \tfrac12{\overbar{\bf N}}\;,\cr}
\eqno(5.1)$$
along with definitions of the symbols involved:
$$\eqalign{P^2 = \sq F^2\left[(w_1-w_2)u^2 +(w_3-w_1)\right]\;,&
\qquad {\overbar P}^2 = \sq F^2\left[(w_1-w_2){\overbar u}^2 +(w_3-w_1)\right]\;,\cr
  p = \log\tan(\theta/2) + i(\psi+\pi/2)\;,&\quad \eqalign{u = &\ \cosh p = -2{\sin\psi +i\cos\psi
  \cos\theta\over
\sin\theta}\;,\cr
{\overbar u} = &\ \cosh {\overbar p} = -2{\sin\psi -i\cos\psi\cos\theta\over
\sin\theta}\;.\cr}}\eqno(5.2)$$
The overbar is used here to indicate complex conjugation in the situation where we treat the variables
$\{\mu,\theta,\psi,\varphi\}$ as real-valued.  Since our goal is in fact to determine general
complex-valued
solutions, that approach is still valid in the more general case, where we note that the
variable $u$ may take
on all values in the complex plane.  The equation that determines the (potential) function
$\Omega$ is then
given by the following, in terms of functions of $\{\mu,\theta,\psi\}$:
$$\eqalign{e^\Omega = &\ 2F^2\Big\{\left[(w_1-w_2)(\sin\psi + i\cos\psi\cos\theta)^2
 + (w_3-w_1)\sin^2\theta\right]\cr & \hskip1in
\left[(w_1-w_2)(\sin\psi - i\cos\psi\cos\theta)^2 + (w_3-w_1)\sin^2\theta\right]\Big\}^{1/2}\cr\cr
&\qquad = 2F^2\sin^2\theta\,\left|(w_1-w_2)u^2-(w_1-w_3)\right| = \sin^2\theta
\left|P\right|^2\;.}\eqno(5.3)$$
We have shown explicitly that this parametrically-determined function $\Omega$ does indeed satisfy
the sDiff(2) Toda equation, as described in \Es{1.2}, where the function ${\bf N}(\mu,p)$ is given in
\Es{3.17} while the general functions $w_i(\mu)$, that determine the dependence of the metric on the
transverse coordinate $\mu$, are given in terms of the conformal factor $F=c_0(\mu+d_0)$ and the
solutions to the Halph\'en problem, $a_i(\mu)$, in \Es{2.5}.  The functions $a_i(\mu)$ themselves
are given in their most general form in \Es{A3.16}, which constitute a (conformal) M\"obius transformation
of the $i\mu$ upper complex half-plane of the simpler form given in \Es{2.4}.
\vs
This is therefore the most general solution of the Pleba\'nski equation that allows SU(2) symmetry.  It is
hoped that this idea of parametric solutions to a very complicated nonlinear partial differential
equation will be a useful one, and will engender ways to determine yet more general solutions, with smaller
symmetry groups.

\vskip15pt\no
{\bf Appendix I:~~Calculations for the curvature of a vacuum Bianchi IX metric}
\vs
We believe it useful to show
how the requirements of anti-self-duality of the conformal tensor, and the vanishing of the Ricci tensor,
are turned into (ordinary) differential equations that must be satisfied by $F$ and the three $w_i$'s.
Since the purpose of the conformal factor is to affect changes in the Ricci tensor, it is simpler to
first determine these constraints by ignoring the function $F$, i.e., to set it just equal to $+1$,
and then afterward use it to perform a conformal
transformation of the metric, to achieve the desired results.  We also note that since we are
interested in the general case of complex-valued manifolds, for the sDiff(2)Toda equation, any particular
form for the signature is not essential. Therefore, we will
follow a fairly standard approach, and set up a tetrad with Riemannian signature, when variables are
considered as real-valued,
for our calculations, with $F=1$:
$$\eqalign{{\bf g}\Big|_{F=1}\equiv &\ (\dom^1)^2+(\dom^2)^2 + (\dom^3)^2 + (\dom^4)^2\;,\cr
 &\dom^4\equiv H\,d\mu\;,\quad
\dom^i \equiv H{\sig_i\over w_{_I}}
\;,\quad i = 1, 2, 3\;,\qquad  H\equiv  \sqrt{w_1w_2w_3}\;,\cr}\eqno(A1.1)$$
where we use an upper-case index along with a lower-case one when we need the same values, but
to indicate that {\bf no sum} on the values is intended.
\vs
To consider self-duality, of 2-forms, we also need a basis set for the vector spaces of 2-forms;
we create the following two sets of triples
$$\eqalign{{\dform E}^k_\pm \equiv \dom^i\wedge\dom^j \pm \dom^k\wedge\dom^4\;,\quad i,j,k
\hbox{~~from $1, 2, 3$, and cyclic,}\cr}\eqno(A1.2)$$
where the ones with the plus sign are a basis for self-dual 2-forms,
and those with the minus sign are anti-self-dual.  We use
the word ``cyclic" with a fairly standard meaning, i.e.,
to mean that the indices $\{i,j,k\}$ should always take distinct
values and in cyclic order, so that they imply each of the three possible sets of values
$1,2,3$,  and $2,3,1$, and $3,1,2$.
\vs
The curvature, and the connection, of the manifold split into separate constituents for the self-dual and
anti-self-dual parts:  the connection for the two parts, which are determined by separate triplets of 1-forms,
along with separate triplets of 2-forms to determine the curvature.  The complete Cartan relations are of
course written as follows:
$$ d\dom^\alpha = \dom^\mu\wedge\gam^\alpha{}_\mu
\;,\quad \OM^\mu{}_\nu =  d\gam^\mu{}_\nu + \gam^\mu{}_\sigma\wedge\gam^\sigma{}_\nu\equiv
\tfrac12R^\mu{}_{\nu\lambda\rho}\,\dom^\lambda\wedge\dom^\rho\;.\eqno(A1.3)$$
However, the separation into self-dual and anti-self-dual parts
splits the connection and curvature into a pair of triplets of forms for each of them:
$$\gee^i_\pm\equiv \gam_{jk}\pm \gam_{i4} \;,\quad
\OM^i_\pm\equiv \OM_{jk}\pm\OM_{i4} = d\gee^i_\pm -\gee^j_\pm\wedge\gee^k_\pm\;,    \eqno(A1.4)$$
where again the upper subscripts are for the self-dual parts and the lower ones for the anti-self-dual
parts.
Inserting our tetrad, along with considerable calculation, gives us the two triplets of connections
in the following explicit form:
$$\eqalign{ \gee^i_\pm = &\ {a_{\pm_{I}}\overwithdelims() H}\,\dom^i\;,\hskip0.75in
 a_{\pm j}+a_{\pm k} \equiv \pm{w_i'\over w_{_I}} + {w_jw_k\over w_i}\;,\cr}\eqno(A1.5)$$
 where the new functions $a_{\pm k}(\mu)$ are defined by the above triplet of equations
in terms of the original $w_i$'s,
and the prime denotes the derivative with respect to $\mu$.
 [Note that within this approach
the distinction between self-dual
and anti-self-dual may be switched simply by switching the sign of the coordinate $\mu$.]
An immediate observation is
 that several of the connection coefficients are zero, since each of the
two triplets of connections is spanned only by one basis vector, so that we have only 6 connection coefficients,
instead of the maximal possible number of 24.  These connections are then used to generate their respective
curvatures, either self-dual or anti-self-dual, which gives us the following:
$$\eqalign{\OM^i_\pm &\ = (R_{jkjk}\pm R_{jki4})\dom^{\scriptscriptstyle J}
\wedge\dom^{\scriptscriptstyle K}
 + (R_{jki4}\pm R_{i4i4})\dom^{\scriptscriptstyle I}\wedge\dom^{\scriptscriptstyle4}
\equiv {\cal Z}_{\pm i}\dom^{\scriptscriptstyle J}
\wedge\dom^{\scriptscriptstyle K} + {\cal K}_{\pm i}\dom^{\scriptscriptstyle I}\wedge\dom^{\scriptscriptstyle4}\cr &
\hskip0.7in = \dfrac1{H^2}\left\{ \left({H^2\over w_{_I}}a_{\pm_i} -
 a_{\pm_J}a_{\pm_K}\right)\,\dom^j\wedge\dom^k -w_{_I}{a_{\pm_I}\overwithdelims()w_{_I}}^\prime\,
\dom^i\wedge\dom^4 \right\}\;.\cr}\eqno(A1.6)$$
Once again,
since each of the $\OM^i_\pm$ are spanned by only two of the basis 2-forms, namely $\dom^j\wedge\dom^k$ and
$\dom^i\wedge\dom^4$, with all other possible coefficients zero, there
appear to be only $3\times3 = 9$ non-zero components of the curvature tensor,
namely $R_{jkjk}$, $R_{jki4}$, and $R_{i4i4}$.  However, one of them is not linearly independent, since
 the first Bianchi identity causes those three of the form $R_{jki4}$ to sum
to zero, i.e., $R_{1234}+R_{2341}+R_{3412} = 0$, so only 8 of these are independent.
There are also constraining
relations between the convenient labels ${\cal Z}_{\pm i}$ and ${\cal K}_{\pm i}$.
 The most important of those are
the following:
$${\cal Z}_{+i}-{\cal Z}_{-i} =  {\cal K}_{+i} + {\cal K}_{-i}\;,\qquad
\sum_{i=1}^3\left({\cal Z}_{+i}-{\cal Z}_{-i}\right) =  0 = \sum_{i=1}^3\left({
\cal K}_{+i}+{\cal K}_{-i}\right)\;.\eqno(A1.7)$$
\vs
To divide these components further, we separate those 8 components into those that
constitute the conformal tensor,
$C_{\mu\nu\lambda\eta}$, and
those that define the Ricci tensor, ${\cal R}_{\mu\nu}$, and its trace, $\cal R$:
$$\eqalign{C_{\mu\nu\lambda\eta} = &R_{\mu\nu\lambda\eta}-\tfrac12\left(g_{\mu\lambda}{\cal R}_{\nu\eta} - g_{\mu\eta}{\cal R}_{\nu\lambda}
+g_{\nu\eta}{\cal R}_{\mu\lambda} - g_{\nu\lambda}{\cal R}_{\mu\eta}\right) + \tfrac16\left(g_{\mu\lambda}g_{\nu\eta}
- g_{\mu\eta}g_{\nu\lambda}\right){\cal R}\;,\cr
&{\cal R}_{\nu\mu} =  {\cal R}_{\mu\nu} = R^\lambda{}_{\mu\lambda\nu}\;,\qquad {\cal R}\equiv R^{\mu\nu}{}_{\mu\nu}\;.
}\eqno(A1.8)$$
It is simpler, and equivalent, to divide the conformal tensor components into their self-dual and anti-self-dual parts
on the basis of their definition as 2-forms, rather than their other pair of indices, using our two basis sets for 2-forms
to accomplish this:
$$\eqalign{ {\dform{\cal C}}_{\pm \mu\nu}&\ =
 {\cal C}_{i\pm\mu\nu}{\dform E}^i_\pm\;,\qquad
 \cdf_{\pm ij} = \cdf_{\pm k4}\;,}\eqno(A1.9)$$
 where the second set of equalities is a generic statement about the symmetries of this tensor, which in the most
 general case only has 5 components for the self-dual part and another 5 for the anti-self-dual part.
 However, in our particular case, using
 the forms for the curvature tensor given in \Es{A1.6} we find that only the following two sets of triplets of
components, ${\cal C}_{i\pm jk}$,
are non-zero, i.e., those where all of $i$, $j$, and $k$ are different.  However, as well the
 sum of each triplet is also zero by the first Bianchi identity, so that we have
only the two independent components for the self-dual part and another two for the anti-self-dual part:
$$\eqalign{2H^2\;{\cal C}_{i\pm jk}&\ = 2\cw{i} - \cw{j}-\cw{k}\;,\cr
\cw{i}&\ \equiv {\cal Z}_{i\pm} \pm {\cal K}_{i\pm} \cr
&\qquad = H^2[R_{jkjk}\pm2R_{jki4}+R_{i4i4}] =
\mp {a_{\pm i}}^\prime + a_{\pm i}(a_{\pm j}+a_{\pm k}) - a_{\pm j}a_{\pm k}\;.}
\eqno(A1.10a)$$
{\bf Once} again only two of the three elements in either one of these two triplets is independent, since it is
straightforward to see that
$$ {\cal C}_{1\pm23} + {\cal C}_{2\pm 31} + {\cal C}_{3\pm 12} = 0\;.\eqno(A1.10b)$$
The Ricci tensor of course involves
components from both the self-dual and the anti-self-dual sides, which can most conveniently be
described in terms of both the $\cw{i}$ and also the additional parts ${\cal K}_{i\pm}$:
$$\eqalign{
2H^2{\cal R}_{44} = &\ \sum_{i=1}^3\left({\cal K}_{+i} - {\cal K}_{-i}\right) = 2\sum_{i=1}^3{\cal K}_{+i}\;,\qquad
\qquad 2H^2{\cal R} =4\sum_{i=1}^3{\cal W}_{+i}\;,\cr
2H^2{\cal R}_{ii} =&\ -(a_{+i}-a_{-i})^\prime + (a_{+j}+a_{-j})(a_{+k}+a_{-k})\cr
&\hskip.5in = 2\left({\cal W}_{+j} + {\cal W}_{+k} + 2{\cal K}_{+i} - H^2{\cal R}_{44}\right)
\;.\cr}\eqno(A1.11)$$
\vs
We intend to require that the conformal curvature be anti-self-dual, which gives us {\bf only} two constraints on the
 three ${\cal W}_{+i}$'s; in
order to obtain a third one, we require as well that the scalar curvature, ${\cal R}$, vanish, which
then requires all three of the ${\cal W}_{+i}$'s to vanish.   This system of three equations, for the
three unknown functions $a_{+i}$, is usually referred to as the
Halphen system,\ftp{21} but also by various other names, including often Darboux and/or Brioschi, since there was
considerable interest in their solution in the late part of the 19th century.  Once we have
 distinct forms for the $a_{+i}$'s then \Es{A1.5} constitute a triplet of first-order differential equations
to determine the form of the three $w_i$'s which must also be solved.
A general solution of these 6 equations has been studied by several authors; we prefer the
particular form of the solution used by Babich
and Korotkin\ftp{18}.  In the case where the Ricci scalar vanishes and
all three of the $a_{+i}$'s are distinct they show
that the difference $w_i-a_{+i}$ is independent of $i$.  Referring to that difference
as $g=g(\mu) = 1/(\mu+q_0)$, with $q_0$ an arbitrary constant, we will now
use that information to greatly simplify the previously-given forms for the traceless part
of the Ricci tensor.
We can use the ode's that we have assumed satisfied to re-write the auxiliary quantities
${\cal K}_{+i}$, in a very simple form:
$$\eqalign{{\cal K}_{+i} \equiv &\ -a'_{+ i} + a_{+ i}{w'_{_I}\over w_{_I}}
 =  a_{+j}a_{+k} - a_{+i}{w_jw_k\over w_{_I}}
 = (w_j-g)(w_k-g)-(w_i-g){w_jw_k\over w_i}\cr &
= g^2-g\left(w_j+w_k-{w_jw_k\over w_i}\right)  = -g(g+w_i'/w_{_I})\;,}\eqno(A1.12)$$
 This then allows simple expressions for the components of the
Ricci tensor:
$$\eqalign{H^2{\cal R}_{44} = &\ \sum_{i=1}^3{\cal K}_{+i} = -g\sum_{i=1}^3(g+w_i'/w_{_I})
 = -g\left(3g+\sum_{i=1}^3{w_i'\over w_{_I}}\right)= -g(3g+2H'/H)\;,\cr
 H^2{\cal R}_{ii} =&\ \left({\cal W}_{+j} + {\cal W}_{+k} + 2{\cal K}_{+i} - H^2{\cal R}_{44}\right)
\cr &\qquad = [{\cal K}_{+i} + g(3g+2H'/H)] = g\{g+2[\log(H/w_i)]'\}\;.  }\eqno(A1.13)$$
\vs
Since the sDiff2(Toda) equation generates metrics which have zero Ricci tensor, and the standard form
of the Bianchi IX metric described above still has non-zero diagonal components of that tensor, we must
now implement the conformal transformation of the metric generated by our function $F=F(\mu)$,
 to arrange for the Ricci tensor to vanish as well.  We consider the re-scaled metric,
and correspondingly re-scaled tetrad basis 1-forms, denoting the re-scaled tensors with a ``hat" over
the relevant symbols:
$$\delta_{\mu\nu}\,{\hat\dom}^\mu\mathop{\otimes}_s{\hat\dom}^\nu =
{\hat{\bf g}} \equiv F^2{\bf g} = F^2\delta_{\mu\nu}\,
\dom^\mu\mathop{\otimes}_s\dom^\nu\;,\Long {\hat\dom}^\mu = F\,\dom^\mu\;.\eqno(A1.14)$$
Such a transformation will leave invariant the tetrad components of the conformal tensor,
${\hat C}^\alpha{}_{
\beta\gamma\delta}$.  However, the general
transformation of the tetrad components
of the Ricci tensor and, separately and usefully, its trace, $\cal R$, is given as follows:
$$\eqalign{{\hat{\cal R}}_{\beta\delta}&\ = {\cal R}_{\beta\delta} + X_{\beta\delta}\;,\hskip0.5in
\hat{{\cal R}} = F^{-2}{\cal R} + {\cal X}\;,\cr
&\qquad X_{\beta\delta} = F\left(2\delta^\alpha_\beta\delta^\zeta_\delta + g_{\beta\delta}g^{\alpha\zeta}\right)
\nabla_\alpha\nabla_\zeta F^{-1} - 3F^2\,g_{\beta\delta}\left[g^{\alpha\zeta}(\nabla_\alpha F^{-1})(\nabla
_\zeta F^{-1})\right]\;,\cr
&\hskip1.5in{\cal X} = -6F^{-3}g^{\beta\delta}\nabla_\beta\nabla_\delta F\;.
}\eqno(A1.15)$$
Since we want to maintain unchanged the current zero value for the Ricci scalar, we see that this is
straightforward provided  the function $F$
is ``harmonic." Now we show that a conformal factor which depends only on $\mu$, therefore not
disturbing our symmetry, is sufficient to annul the Ricci tensor as is desired.  Such a dependence
of course simplifies the equations greatly, giving us the following:
$$\eqalign{{\cal X}_{44} = -{1\over H^2}&\,\left[3{F''\over F}
 - 3{F'\overwithdelims()F}^2 - 2{F'H'\over FH}\right]\;,\qquad
{\cal X}_{4i} =  0\;,\cr
{\cal X}_{ij} =  -{1\over H^2}\delta_{ij}&\,\left\{{F''\over F} +
{F'\overwithdelims()F}^2 + 2{F'\over F}[\log(H/w_{_I})]^\prime
\right\}\;,\qquad
{\cal X} =  -6{F''\over H^2F}\;.\cr}\eqno(A1.16)$$
\vs
In this case the requirement that $F$ be harmonic simply reduces to the requirement that it be
 a linear function of $\mu$:
$$ 0 = {\cal X} \propto F'' \quad\Longrightarrow\quad F = c\mu + d\;,\eqno(A1.17)$$
where $c$ and $d$ are constants.
We next go to the earlier-determined, non-zero forms for the Ricci tensor components themselves, from \Es{A1.13},
 to see if this form for $F$ will allow the transformed (trace-free) Ricci tensor to vanish.  Beginning with
 ${\cal R}_{44}$,
where we are now including the requirement that $F'' = 0$, we have
$$H^2F^2{\hat{\cal R}}_{44} = H^2{\cal R}_{44} + H^2X_{44} = -g(3g+2H'/H) +(F'/F)[3F'/F + 2H'/H]\;.\eqno(A1.18)$$
For this to vanish we need only to require that $F'/F = g$:
$$ F = c_0(\mu+d_0)\quad\Longrightarrow\quad F'/F =  {1\over \mu+d_0}\;,
\qquad\hbox{but~~} g = 1/(\mu+q_0)\;.\eqno(A1.19)$$
  Therefore, by setting the two previously arbitrary integration constants, $q_0$ and $d_0$ equal to each other,
  we have accomplished what was desired for ${\cal R}_{44}$.  Since everything is now determined, we must
  now hope that this will also allow
 the remainder of the transformed components of the Ricci tensor to vanish:
$$\eqalign{H^2F^2{\hat{\cal R}}_{ii} =&\  H^2{\cal R}_{ii} + H^2X_{ii}\cr &
 = g(g+2H'/H - 2w_i'/w_{_I}) - (F'/F)[(F'/F) + 2H'/H
- 2w_i'/w_{_I}]\;.}\eqno(A1.20)$$
We see that, yes, once again the choice that $F'/F=g$, is sufficient to
cause these components to vanish as well, which tells us that the metric ${\hat{\bf g}} = (c_0/g)^2{\bf g}$
should also be capable of being generated by a solution to the Pleba\'nski equation,
where $\bf g$ is the Bianchi IX metric described in \Es{2.1}.
\vs\no
{\bf Appendix II:~~The (3-dimensional) Hodge dual in the spherical coordinates}
\vs
We first recall that the Levi-Civita tensor, in 3 dimensions and with an arbitrary metric $g$, with
Riemannian signature, is given by
$$\eta_{\alpha_1\alpha_2\alpha_3} = \sqrt{g}\,\epsilon[\alpha_1,\alpha_2,\alpha_3]\;,\qquad g \equiv
\det(g_{ab})\qquad \eta^{\alpha_1\alpha_2\alpha_3} =
{1\over \sqrt{g}}\,\epsilon[\alpha_1,\alpha_2,\alpha_3]\;,\eqno(A2.1)$$
where the $\epsilon$-symbol is the usual Levi-Civita alternating symbol that takes on only the values
$\pm 1$ and $0$.  Then the general form of the dual of a 1-form or a 2-form, under our metric $\bgamma$, is given by
$$\eqalign{ {\dform{\alpha}} = \tfrac12\alpha_{ab}\,\dom^a\wedge\dom^b\quad
\Longleftrightarrow\quad *\left(\dform{\alpha}\right) = \tfrac12\bgamma^{ac}\bgamma^{bd}\alpha_{ab}\,
\eta_{cdf}\,\dom^f\;,\cr
{\dform{\beta}} = \beta_{a}\,\dom^a\quad
\Longleftrightarrow\quad *\left(\dform{\beta}\right) = \tfrac12\bgamma^{ac}\beta_{a}\,
\eta_{cdf}\,\dom^d\wedge
\dom^f\;,\cr}\eqno(A2.2)$$
Our metric is given explicitly in \Es{3.4}, but is perhaps best here presented in the following symbolic form:
$$\bgamma = \left(\left(\bgamma_{ij}\right)\right)
 \rep \bordermatrix{&\mu&\theta&\psi\cr \mu&\gamma_{\mu\mu}&0&0\cr \theta&0&\gamma_{\theta\theta}&\gamma_{\theta\psi}\cr
\psi&0&\gamma_{\theta\psi}&\gamma_{\psi\psi}}\;,\eqno(A2.3a)$$
along with its determinant:
$$\det\bgamma = \gamma_{\mu\mu}\left(\gamma_{\theta\theta}\gamma_{\psi\psi} - \gamma_{\theta\psi}^2\right)
= \gamma_{\mu\mu}(F^4\sin^2\theta\,\gamma_{\mu\mu}) = (\gamma_{\mu\mu}\,F^2\sin\theta)^2 = (M\,F^6\sin\theta)^2
\;,\eqno(A2.3b)$$
where the value of $M$ is noted in \E{3.2} and in \E{3.4} it is noted that $\gamma_{\mu\mu} = MF^4\;.$
\vs
The simple form of this makes it very easy to present the inverse metric:
 $$\bgamma^{-1} = \left(\left(\bgamma^{ij}\right)\right) \rep {\ell^2\over \gamma_{\mu\mu}}
 \pmatrix{\ell^{-2}&0&0\cr 0 &\gamma_{\psi\psi}& -\gamma_{\theta\psi}\cr
 0& -\gamma_{\theta\psi}&\gamma_{\theta\theta}}\;,\qquad\ell\equiv {1\over F^2\sin\theta}\;,\;.\eqno(A2.3c)$$
We then begin by calculating the duals of the three basis 2-forms and
also, reciprocally, the three basis 1-forms:
$$\left.\eqalign{*(d\theta\wedge d\psi) = &\ \ell\,d\mu\;,\cr
*(d\mu\wedge d\theta) = &\ \ell\left[\gamma_{\theta\psi}d\theta + \gamma_{\psi\psi}d\psi\right]
/\gamma_{\mu\mu}\;,\cr
*(d\mu\wedge d\psi) = &\ -\ell\left[\gamma_{\theta\theta} d\theta + \gamma_{\theta\psi}d\psi\right]
/\gamma_{\mu\mu}\;,\cr}\right\}\quad\left\{\eqalign{
*d\mu = &\ \ell^{-1}d\theta\wedge d\psi\;,\cr
*d\theta = &\ \ell\left[\gamma_{\psi\psi}\,d\psi\wedge d\mu - \gamma_{\theta\psi}\,d\mu\wedge d\theta
\right]\;,\cr
*d\psi = &\ \ell\left[\gamma_{\theta\theta}\,d\mu\wedge d\theta - \gamma_{\theta\psi}d\psi\wedge d\mu\right]
\;,\cr}\right.\eqno(A2.4)$$
where the various coefficients of the 3-metric, $\bgamma$, may be found in \Es{3.4}.
\vs
It is then straightforward to write down the duals of an arbitrary 1-form $\dform\alpha$
 and 2-form $\dform\beta$:
$$\eqalign{ {\dform\alpha} \equiv & \ H_1\,d\mu + H_2\,d\theta + H_3\,d\psi\;,\quad {\dform\beta} =
J_1\,d\theta\wedge d\psi + J_2\,d\mu\wedge d\theta + J_3\,d\mu\wedge d\psi\cr
&\hskip1.5in\Longleftrightarrow\cr
*({\dform\alpha}) = & \ell^{-1}H_1d\theta\wedge d\psi + \ell(\gamma_{\theta\theta}H_3
 - \gamma_{\theta\psi}H_2)
d\mu\wedge d\theta - \ell(\gamma_{\psi\psi} H_2 - \gamma_{\theta\psi}H_3)d\mu\wedge d\psi\;,\cr
*({\dform\beta}) = & \ell J_1 d\mu + {\ell\over \gamma_{\mu\mu}}\left[\gamma_{\theta\psi}J_2
-\gamma_{\theta\theta}J_3\right]\,d\theta + {\ell\over \gamma_{\mu\mu}}\left[\gamma_{\psi\psi}J_2
- \gamma_{\theta\psi}J_3\right]\,d\psi\;.\cr}\eqno(A2.5)$$
Using \Es{3.2-3} for $1/V$ and $\dform{\omega}$,  as well as the expressions for the coefficients of
the metric $\bgamma$, in the expression \E{3.7}, we obtain the expression for $ds$ as given in
\E{3.8} in the main text.
\vs\no
{\bf Appendix III:~~Integration for $q=q(\mu,\theta,p)$, and general Theta Functions}
\vs
We intend here to show that the solution of the three (compatible) differential equations for $q$ that are
described in \Es{3.15} is in fact that given in \E{3.16}, i.e.,
$$\eqalign{  2q = \tfrac12P\cos\theta + {\bf N}(\mu,p)\;,&\qquad P\sin\theta\,(2\,dq) =
\Ep = {\cal A}\,d\mu + {\cal B}\,d\theta + {\cal C}\,dp\;,\cr
&\hskip-.5in P^2 \equiv \sq F^2[(w_1-w_2)\cosh^2p - (w_1-w_3)]\;.}\eqno(A3.1)$$
We begin by inserting just the first term of the form for $2q$ into the equation for $\Ep$.  All
the dependence on the variables is given explicitly in this case, and we obtain:
$$\eqalign{P\sin\theta\,& d(\tfrac12P\cos\theta) =  -\tfrac12P^2\sin^2\theta\,d\theta + \tfrac14\sin\theta\cos\theta\,d(P^2)\cr
& = -\tfrac12e^{2f}\,d\theta + F^2\sin\theta\cos\theta(w_1-w_2)\cosh p\sinh p\;dp + \tfrac14\sin\theta\cos\theta\,
{\p P^2\over \p\mu}\,d\mu \cr & \quad =  -\tfrac12e^{2f}\,d\theta + F^2\sin\theta\cos\theta(w_1-w_2)u\sqrt{u^2-1\,}\,dp\cr
&\hskip.7in +
F^2\sin\theta\cos\theta[w_3(w_1-w_2)u^2+w_2(w_3-w_1)]\,d\mu\;.}\eqno(A3.2)$$
These do in fact satisfy exactly all the $d\theta$ terms in $\Ep$ and all those other terms in $\Ep$ that
have an explicit $\cos\theta$.  Therefore we are left with the two remaining equations to determine the
yet-unknown function ${\bf N} = {\bf N}(\mu,p)$:
$$\eqalign{P{\p {\bf N}\over \p\mu} =&\ F^2w_3(w_1-w_2)\cosh p\sinh p\;,\cr
P{\p {\bf N}\over \p p} = &\ F^2[(w_1-w_2)\cosh^2p - w_1]\;.\cr}\eqno(A3.3)$$
We begin this part of the problem by first determining an integral for the
equation involving $\p {\bf N}/\p p$:
\par\no
$$\eqalign{\sqrt{2}\,{\bf N}  = &\  F\int dp\,\sqrt{(w_1-w_2)\cosh^2p-(w_1-w_3)}\cr
&\hskip.6in - F\int dp\,
{w_3\over \sqrt{(w_1-w_2)\cosh^2p-(w_1-w_3)}}\cr
&  = \sqrt{w_1-w_3}F\int dp\,\sqrt{k^2\cosh^2p - 1} - {w_3\over \sqrt{w_1-w_3}}F\int {dp\over
\sqrt{k^2\cosh^2p-1}}\cr
&\hskip.5in = F\left[\sqrt{(w_1-w_3)}E(u,k) + {w_3\over \sqrt{w_1-w_3}}
F(u,k)\right]
\;,\cr}\eqno(A3.4)$$
where $F(u,k)$ and $E(u,k)$ are the standard first and second incomplete elliptic integrals,
and  $(w_1-w_2)/(w_1-w_3) = k^2$ is the Jacobi modulus for Jacobi elliptic
functions, as will be shown below:
$$\eqalign{ F(z,k) \equiv &\int_0^z{da\over \sqrt{(1-a^2)(1-k^2a^2)\,}}
 = \int_0^{\sin^{-1}z}\hskip-0.2in{d\theta\over
\sqrt{1-k^2\sin^2\theta\,}} = \int_0^{\sn^{-1}(z,k)}\hskip-.15in dw\;,\cr
E(z,k) \equiv &\int_0^zda{\sqrt{1-k^2a^2}\over \sqrt{1-a^2}} = \int_0^{\sin^{-1}z}\hskip-.15in d\theta\,
{\sqrt{1-k^2\sin^2\theta}}
= \int_0^{F(z,k)}\hskip-.15in dw \,\dn^2(w,k)\;.\cr}\eqno(A3.5)$$
\vs
Although this is indeed the desired solution, it may be put into much more reasonable forms provided
we now make explicit use of our solutions for the functions $w_i$, and also the $a_i$,
as noted in \Es{2.4} and (2.5),
in terms of $F = c_0(\mu+d_0)$ and the theta coefficients.  However,
those coefficients may also be expressed in terms of complete elliptic integrals:\ftp{23}
$$\eqalign{\pi a_3(\mu) = &\ 2\pi i{d\over d\mu}\log\vartheta_2(0\mid i\mu) = -2\,K(k)E(k) \;,\cr
 \pi a_2(\mu) =&\ 2\pi i{d\over d\mu}\log\vartheta_3(0\mid i\mu) = -2\,K(k)[E(k) - k'^2K(k)]\;,\cr
\pi a_1(\mu) = &\ 2\pi i{d\over d\mu}\log\vartheta_4(0\mid i\mu) = -2\,K(k)[E(k)-K(k)]\;,\cr
&w_i(\mu) = a_i(\mu) +  {d\over d\mu}\log F(\mu) = a_i(\mu) + {1\over \mu+ d_0}\;, \cr
k'^2 \equiv 1-k^2\;,&\qquad \mu = -i\tau = K'(k)/K(k)\;,\cr}\eqno(A3.6)$$
where the extra factor of $\pi$ appears because of our normalization, following Hancock,\ftp{23}
for the arguments of the theta functions.  Normalizations vary considerably from author to author,
concerning the arguments of these functions.  We will present ours at the end of this section.
\vs
This allows us to present the two differences of functions $w_i$ that appear in
our integral in a different way, more easily showing the values
for the integration being performed for $\bf N$:
$$ w_1-w_3 = \dfrac2\pi K^2(k)\;,\qquad w_1-w_2 = \dfrac2\pi k^2K^2(k)\;,\eqno(A3.7)$$
which allows for the following re-presentation of the result for $\bf N$:
$$\eqalign{
\sqrt{\pi}\,{\bf N}= c_0(\mu+d_0)\left[K(k)E(u,k)-E(k)F(u,k)\right] + \tfrac\pi2c_0{F(u,k)\over K(k)}\;,}
\eqno(A3.8)$$
It is of course true that there might also be some ``constant of integration," which would depend
on $\mu$.  We show that no such constant is needed by inserting this value for $\bf N$ back into the differential
equation involving its derivative with respect to $\mu$, and finding that it gives exactly
the desired right-hand side; i.e., the pde is satisfied exactly with the value given above.
\vs
We would like, however, to present this result also in some other formats, where the dependence on $\mu$ is made
more explicit.  The simplest next step is to turn it into a form involving the Jacobi Zeta function,
$Z(w,k)$,\ftp{23, 24}
and then use its relationship to the theta functions with both arguments non-zero.
\par\no
$$\eqalign{Z(w,k) = &\ E(w,k) - {E(k)\over K(k)}w\;,\cr
K(k) Z[F(u,k),k] = &\ K(k)E[F(u,k),k] - E(k)F(u,k)\;,\cr
K(k) Z[2K(k)z;k] = &\ \tfrac12{d\over dz}\log\,\vartheta_4(z\mid i\mu)\;,\cr
K(k)Z(a,k) = &\  \Pi_1[1,a,k] \equiv \Pi_1(a,k)\;,}\eqno(A3.9)$$
where the last line shows the relationship between the Jacobi Zeta function and the complete elliptic integral
of the third kind, in the form originally given by Jacobi, and used by Whittaker and Watson.\ftp{24}  When the
first argument is $1$ the integral is referred to as {\it complete}, and the value of that argument is often
not shown; however, especially with $\Pi_1$, the first argument is often\ftp{24} given in terms of $F(z,k)$ instead of
$z$ as has been done here, so that then that argument would be $K(k)$.  The more usual form of the integral of the third kind is the form due to Legendre, $\Pi(z,\alpha^2,k)$,
which was used by Olivier,\ftp{16} in his integration for the coordinates $q$, and $\bq$, in the special case when
our conformal factor $F$ is simply a constant.  The Legendre form is
related to the Jacobi form as follows, where the coefficients are various Jacobi elliptic functions,\ftp{24}
and we also give their basic definitions, as integrals:
$$\eqalign{
\Pi(z,\alpha^2,k) \equiv &\int_0^z{dt\over (1-\alpha^2t^2)\sqrt{(1-t^2)(1-k^2t^2)}} = \int_0^{F(z,k)}
{dv\over 1-\alpha^2\sn^2(v,k)}\;,\cr
\Pi_1(z,a,k) \equiv &\ k^2\sn(a,k)\cn(a,k)\dn(a,k)\int_0^{F(z,k)}dv\,{\sn^2(v,k)\over 1-k^2\sn^2(a,k)\sn^2(v,k)}\;,
\;,\cr
\Pi_1[z,a,k] = &\ {\cn (a,k) \dn(a,k)\over \sn(a,k)}\left\{\Pi[z,k^2\sn^2(a,k),k]-z\right\}\;.\cr}
\eqno(A3.10)$$

This allows us to re-write our desired function $\bf N$ in several different, equivalent ways, where we choose
the following one as most useful for our purposes:
$$\eqalign{ \sqrt{\pi}\,{\bf N}(\mu,p) =
 c_0\left[\pi z + \tfrac12(\mu+d_0){d\over dz}\log\vartheta_4(z\mid i\mu)\right]
 &\Big|_{z=\sfrac12F(\cosh p,k)/K(k)}\;,\cr}\eqno(A3.11)$$
 where we recall here the relation of $p$ and $u$ to the original spherical coordinates, $\theta$ and $\psi$,
 $$ p \equiv \log\tan(\theta/2) + i\psi \;,\qquad u \equiv
 \cosh p =  -{\sin\psi + i\cos\psi\cos\theta\over \sin\theta}\;,\eqno(A3.12)$$
 and, in terms of these variables, our important quantity $P^2(\mu,p)$ is given by
 $$ P^2(\mu,p) = {e^{2f}\over \sin^2\theta} = -\tfrac1{\pi}\left[2c_0(\mu+d_0)K(k)\dn(a,k)\right]^2\;.\eqno(A3.13)$$
\vs
Completing our picture we put here the complete definitions for
the theta functions, which are everywhere analytic functions of their first argument, $z$, periodic
with period 1, while they are
analytic in the upper half plane for their second argument, $\tau$.  As well, they have power series
expressions which converge very fast and are generators of the usual Jacobi elliptic functions in that
those functions are ratios of the theta functions:
$$\eqalign{\vartheta_4(z|\tau)
 =\ &1+2\sum_{n=1}^{+\infty}(-1)^n\,q^{n^2}\,\cos(2\pi nz) \;,\cr
 \vartheta_3(z|\tau) = \ &1+2\!\sum_{n=1}^{+\infty}
q^{n^2}\cos(2\pi nz)\,,\cr
\vartheta_2(z|\tau)
= &\quad2\sum_{m=0}^{+\infty}q^{(m+\sfrac12)^2}\,\cos[(2m+1)\pi z]
\;,\cr
\vartheta_1(z|\tau)
= &\quad2\sum_{m=0}^{+\infty}(-1)^mq^{(m+\sfrac12)^2}\,\sin[(2m+1)\pi z]\cr
&\hskip1.5in q\equiv e^{i\pi\tau} = e^{-\pi\mu}\;,}\eqno(A3.14)$$
and we note again that there are various other normalizations for the arguments of these functions,
often including the factor $\pi$ into the argument,\ftp{24} so that, then, they have period $\pi$.
\vs
A last thing to do here is to provide more detail as to how one acquires the other 3 parameters for
the general solution\ftp{18} to the Halphen problem, via a M\"obius transformation of the
$\tau$ (upper) half-plane, accompanied by appropriate transformations of
the dependent functions, where in this brief section we use the ``overbar," as in $\overbar\tau$,
to indicate the result after the transformation, rather than it having any relation to the complex
conjugation operation:
$$\left.\eqalign{\tau\ \longrightarrow\ &{\overbar \tau} \equiv {a\tau+b\over c\tau+d}\;,\cr
a_{+i}\ \longrightarrow\ &{\overbar a}_{+i}({\overbar \tau})\equiv (c\tau+d)^2a_{+i}[\tau({\overbar \tau})]
 + c(c\tau+d)\;,\cr
 w_i\ \longrightarrow\ &{\overbar w}_i({\overbar \tau}) \equiv (c\tau+d)^2w_i[\tau({\overbar \tau})]\;,\cr}
\right\}\;; \qquad ad-bc = +1\;.\eqno(A3.15)$$
Therefore, when we include these 3 parameters, the general solution has the following form, where we retain
the overbars:
$$\eqalign{ {\overbar a}_{+i}({\overbar \tau})&\ = 2{d\over d{\overbar \tau}}\log\vt_{5-i}\left({d{\overbar \tau}-b\over a-c{\overbar \tau}}
\right) + {c\over a-c{\overbar \tau}} \cr
{\overbar w}_i({\overbar \tau}) = &\ {(c\tau+d)^2\over \tau+q_0} + {\overbar a}_{+i}({\overbar \tau}) - c(c\tau+d)
 = {\overbar a}_{+i}({\overbar \tau}) + {1\over
{\overbar \tau}+{\overbar q}_0}\;;\quad {\overbar q}_0 = {aq_0-b\over d-cq_0}\;.}\eqno(A3.16)$$
If one makes the particular, allowed choice of the 3 parameters in the transformation, of
$d=a=1$ and $c=0=b$, then this more general form reduces to our earlier, particular form,
as expected.

\vskip15pt

\leftline{{\bf References:}}\par
\parindent=0pt
\baselineskip=12pt
\def\hi{\hangindent=20truept}
\def\bn#1{{\bf #1}}

\frenchspacing
\smaller

\hi ~1.  J.F. Pleba\'nski, ``Some solutions of complex Einstein equations,"
J. Math. Phys. \bn{12}, 2395-2402 (1975).

\hi ~2.  The choice of anti-self-dual rather than self-dual is simply following earlier history.  The
difference is simply a choice of orientation, or, from the point of view of the (nonlinear) pde's
 to be discussed in this paper, it is simply an overall sign change of the independent variable on which
 the important functions depend.

\hi ~3. F. Neyzi, M.B. Sheftel, and D. Yazici, ``Symmetries, Integrals, and Three-Dimensional Reductions
of Pleba\'nski's Second Heavenly Equation," Physics of Atomic Nuclei \bn{70}, 584-592 (2007);
E.V. Ferapontov and M.V. Pavlov, ``Hydrodynamic reductions of the heavenly equation," Cl. Qu. Grav. \bn{20},
2429-2441 (2003).  Generalizations of the Pleba\'nski equation to hyperheavenly spaces are also still
awaiting much more effort.  A recent pair of papers on that subject are A. Chudecki and M. Przanowski,
``From hyperheavenly spaces to Walker and Osserman spaces: I", and also part II, in
Class. Quantum Grav. {\bf 25} 145010 (2008) and  235019 (2008).

\hi ~4.  K.P. Tod and R.S. Ward, ``Self-dual metrics with self-dual Killing
vectors," Proc. R. Soc. London {\bf A368}, 411-427 (1979).  See also, for instance,
G.W. Gibbons and Malcolm J. Perry, ``New gravitational instantons and
their interactions," Phys. Rev. D {\bf 22}, 313-321 (1980).

\hi ~5.  C.P. Boyer and J.D. Finley, III, ``Killing vectors in self-dual, Euclidean
Einstein spaces," J. Math. Phys. \bn{23}, 1126-1130 (1982).  See also J.D. Finley, III and J.F.
Pleba\'nski, ``The classification of all \hsp ~spaces admitting a Killing
vector," J. Math. Phys. \bn{20}, 1938-1945 (1979).  The equation is also often referred to as
the ``Boyer-Finley" equation.

\hi ~6. The name we use was first used by Kanehisa Takasaki and T. Takebe,
``SDiff(2) Toda Equation---Hierarchy, Tau Function, and symmetries,"
Lett. Math. Phys. \bn{23}, 205-214 (1991).

\hi ~7.  In earlier work on this equation, the metric was still presented in the format
shown in \Es{1.1}, in the original Pleba\'nski variables.
We believe the first appearance of the reduction
to this form was given in Daniel Finley and John K. McIver, ``Generalized Symmetries for the SDiff(2)Toda
Equation," published in {\it Topics in Mathematical Physics,
General Relativity and Cosmology, in Honor of
 Jerzy Pleba\'nski}, H. Garc\'\i a, B. Mielnik, M. Montesinos \& M. Przanowski, Eds.,
  World Scientific, London, 2006, p. 177-191.

\hi ~8. A small sampling is given by the following:
 D.M.J. Calderbank
and Paul Tod, ``Einstein Metrics, Hypercomplex
Structures and the Toda Field Equation," Diff. Geom. Appl. {\bf 14}, 199 (2001);
Y. Nutku and M.B. Sheftel, ``A family of heavenly metrics," arXiv:gr-qc/0105088v4 (2002);
E.V. Ferapontov, D.A. Korotkin and V.A. Schramchenko,
``Boyer-Finley equation and systems of hydrodynamic type," Cl. Qu.
Grav. {\bf 19}, L205-L210 (2002);
Manuel Ma\~nas and Luis Mart\'\i nez Alonso, ``A
hodograph transformation which applies to the Boyer-Finley equation,"
Phys. Lett. A, {\bf 320}, 383-388 (2004).

\hi ~9.  L. Martina, M.B. Sheftel and P. Winternitz, ``Group foliation and
non-invariant solutions of the heavenly equation,"
J. Phys. A {\bf 34}, 9423 (2001).  M.B. Sheftel, ``Method of group foliation,
hodograph transformation and non-invariant solutions of the Boyer-Finley
equation," arXiv:math-ph/0305040v1.

\hi 10.
R.M. Kashaev, M.V. Saveliev, S.A. Savelieva, and A.M. Vershik, ``On
nonlinear equations associated with Lie algebras of diffeomorphism groups
of two-dimensional manifolds," in {\it Ideas and Methods in Mathematical
Analysis, Stochastics, and Applications,} S. Albeverio, J.E. Fenstad, H.
Holden and T. Lindstrom (Eds.) Vol. 1, Cambridge University Press, Cambridge,
UK, 1992, p. 295 ff.

\hi 11.
 M.V. Saveliev and A.M. Vershik, ``New Examples of Continuum Graded
Lie Algebras," Phys. Lett. \bn{A143}, 121-128 (1990).

\hi 12. R. Hern\'andez H., D. Levi, M.A. Rodriguez, \& P. Winternitz, ``Relation between
B\"acklund transformations and higher continuous symmetries of the Toda equation," J. Phys.
A {\bf 34}, 2459-2465 (2001);
J. D. Finley, III \& John K. McIver, ``Non-Abelian
Infinite Algebra of Generalized Symmetries for the SDiff(2)Toda
Equation,"   J. Phys. A. {\bf 37}, 5825-5847 (2005).

\hi 13. Some selected papers on the approach via Bianchi IX symmetries are the following:
V.A. Belinskii, G.W. Gibbons, D.W. Page, and C.N. Pope,
``Asymptotically Euclidean Bianchi IX metrics in quantum gravity," Phys. Lett. \bn{76B}, 433-5 (1978);
R.S. Ward, ``Einstein-Weyl spaces and SU($\infty$) Toda fields,"  Class. Qu. Grav. \bn7, L95-L98 (1990).

\hi 14. S. Chakravarty, ``A class of integrable
conformally self-dual metrics," Cl. Qu. Grav., {\bf 11}, L1-L6 (1994), and
M.J. Ablowitz, S. Chakravarty, and R. Halburd, ``The Generalized Chazy Equation and
Schwarzian Triangle Functions," Asian J. Math. {\bf 2}, 619-624 (1998).

\hi 15. N.J. Hitchin, ``Twistor Spaces, Einstein Metrics and Isomonodromic
Deformations," J. Diff. Geom. {\bf 42}, 30-112 (1995).
 S. Okumura, ``The indefinite anti-self-dual metrics and the Painlev\'e equations,"
J. Math. Phys. {\bf 44}, 4828-4838 (2003).

\hi 16. D. Olivier, ``Complex Coordinates and K\"ahler Potential for the Atiyah-Hitchin Metric,"
Gen. Rel. Grav. {\bf 23}, 1349-1362 (1991).

\hi 17. K.P. Tod, ``Scalar-flat K\"ahler and hyper-K\"ahler metrics from Painlev\'e III," Cl. Qu. Grav. \bn{12},
1535-1547 (1995).  Consider also our Ref. 25.

\hi 18.  M.V. Babich and D.A. Korotkin, ``Self-Dual SU(2)-Invariant Einstein Metrics and Modular Dependence of
Theta Functions," Lett. Math. Phys. \bn{46}, 323-337 (1998).

\hi 19. Radu A. Iona\c s, ``Elliptic constructions of hyperk\"ahler metrics I:  The Atiyah-Hitchin manifold,"
    arXiv:0712.3598v1 [math.DG].

\hi 20.  Material related to the use of theta functions in these contexts is contained in various papers
of Yousuke Ohyama, including ``Differential Relations of Theta Functions," Osaka J. Math.
{\bf 32}, 431-450 (1995).

\hi 21.  ``Sur un syst\`eme d'\'equations diff\'erentielles," G. Halphen, C.R. Acad. Sci. Paris \bn{92}, 1101-3 (1881).
\hi 2. ``Sur certains syst\`emes d'\'equations diff\'erentielles," G. Halphen, C.R. Acad. Sci. Paris \bn{92},
1404-6 (1881).

\hi 22.  Michael Atiyah, ``Low-energy scattering of
non-Abelian monopoles," Phys. Let. A {\bf 107}, 21-25 (1985), and M.F. Atiyah \& N.J. Hitchin, ``The
geometry and dynamics of magnetic monopoles," Princeton Univ. Press, NJ, 1988, as well as Ref. 16.

\hi 23.  Harris Hancock, {\it Lectures on the Theory of Elliptic Functions,} Dover Pub., New York, 1958.  I
note that at least two other references are good and easily available for elliptic integrals and elliptic
functions; however, they have different conventions about the labels for the arguments. The first of these is
Paul F. Byrd and Morris D. Friedman, {\it Handbook of Elliptic Integrals for Engineers and Scientists,}
2nd Edition, Revised, Springer-Verlag, New York, 1971, while the second is the next reference.

\hi 24.  E.T. Whittaker and G.N. Watson, {\it A Course of Modern Analysis,} Cambridge Univ. Press, 1950.

\hi 25. K.P. Tod, {\it Self-dual Einstein metrics from the Painlev\'e equation,}
 Phys. Lett. {\bn A190}, 221-224 (1994).

\bye